\documentclass[a4,pra,aps,epsfig,preprint]{revtex4}
\usepackage{amsmath}
\usepackage{amssymb}
\usepackage{epsfig}
\usepackage{subfigure}
\usepackage{ulem}
\usepackage{mathrsfs}
\usepackage{txfonts}
\usepackage{bbm}
\newcommand{\tql}{\textquotedblleft}
\newcommand{\tqr}{\textquotedblright}

\begin{document}
\title{Experimental NMR implementation of a robust quantum search algorithm}
\author{Avik Mitra$^{\rm a}$, Avatar Tulsi$^{\rm b}$ and Anil Kumar$^{\rm a}$\footnote[2]{Raja Ramanna Fellow, email: 
anilnmr@physics.iisc.ernet.in}}
\affiliation{\scriptsize $^{\rm a}$NMR Quantum Computation and Quantum Information Group, Department of Physics and NMR Research Centre, Indian Institute
of Science, Bangalore- 560012\\ $^{\rm b}$ Department of Physics, Indian Institute of Science, Bangalore, 560012.} 
\begin{abstract}
Grover's quantum search algorithm, involving a large number of qubits, is highly sensitive to errors in the physical implementation of the unitary operators. This poses an intrinsic limitation to the size of the database that can be practically searched. The lack of robustness of Grover's algorithm for a large number of qubits is due to quite stringent `{\it phase-matching}' condition. To overcome this limitation, Tulsi suggested a modified search algorithm [PRA 78, 022332] which succeeds as long as the errors are reproducible and reversible while Grover's algorithm fails. Such systematic errors arise often from imperfections in apparatus setup e.g. the errors arising from imperfect pulse calibration and offset effect in NMR systems. In this paper, we report the experimental NMR implementation of the modified search algorithm and its comparison with the original Grover's algorithm. We experimentally validate the theoretical predictions made by Tulsi.
\end{abstract}
\maketitle

\section{Introduction}

\indent Quantum computation has developed as an exciting field of research in the last decade and it has generated wide interest among scientists and engineers. It offers the opportunity of creation of algorithms that are radically different and more efficient as compared to their classical counterparts. Shor's prime factorization algorithm~\cite{shor} and Grover's quantum search algorithm~\cite{grov} have theoretically demonstrated the power of quantum algorithms. However, the experimental implementation of the quantum algorithms is still quite challenging. Nuclear Magnetic Resonance (NMR) has been the vanguard among the presently available techniques for physical implementation of quantum algorithms~\cite{nmrqc,grochu}. Till date, the algorithms have been tested on systems with a small number of qubits with a presumption that once a quantum computer with large number of qubits are made, more real world application of the algorithms can be implemented. Implementation of the quantum algorithms on very large system requires the application of a large number of unitary operators. As any physical implementation involves some amount of error which accrue when the unitary operators are applied in tandem, physical implementation of an algorithm in a large system becomes difficult. The sensitivity of the algorithm to small errors can lead to it's failure.

Grover's quantum search algorithm, or more generally the quantum amplitude amplification algorithm, is designed to search a marked item from an unsorted database~\cite{grov,qaa}. It drives a quantum computer from a prepared initial state $|s\rangle$ to a desired target state $|t\rangle$, which encodes the marked item. Generally, $|s\rangle$ is prepared by applying a unitary operator $U$ on a particular basis state $|0\rangle$, i.e. $|s\rangle = U|0\rangle$. The heart of the algorithm is the Grover's iteration operator $\mathcal{G}$ given by
\begin{eqnarray}
\mathcal{G} = I_{s}I_{t} = UI_{0}U^{\dagger}I_{t}\ ,\ \ \ I_{\psi} = \mathbbm{1}-2|\psi\rangle\langle \psi|. \label{grop}
\end{eqnarray}
Thus $I_{\psi}$ is the selective phase inversion of the state $|\psi\rangle$. If $\alpha = |\langle t|s\rangle|$ then $\pi/4\alpha$ times iteration of $\mathcal{G}$ on $|s\rangle$ yield the target state $|t\rangle$ with a high probability. For searching a database of $N$ items, the initial state $|s\rangle$ is chosen to be the equal superposition of all basis states each of which has a probability amplitude $1/\sqrt{N}$. It is generated by applying the Walsh-Hadamard transform $W$ on the basis state $|0\rangle$, i.e. $|s\rangle = W|0\rangle$. Since $|t\rangle$ is a unique basis state, $\alpha = 1/\sqrt{N}$ and $O(\sqrt{N})$ times iterations of $\mathcal{G} = WI_{0}WI_{t}$ on $|s\rangle$ yield the target state $|t\rangle$. 

In this paper, we consider the case when the implementation errors cause the deviations in selective phase inversions, $I_{s}$ and $I_{t}$. In other words, we want the apparatus to implement $\{I_{s},I_{t}\}$ but due to errors, the apparatus implements $\{I_{s}^{\phi},I_{t}^{\varphi}\}$ where 
\begin{equation}
I_{\psi}^{\omega} = \mathbbm{1}-(1-e^{\imath \omega})|\psi\rangle\langle \psi|
\end{equation}
is the selective phase rotation of $|\psi\rangle$ by angle $\omega$. Then the Grover's operator becomes $\mathcal{G} = I_{s}^{\phi}I_{t}^{\varphi}$ and the well-known \emph{phase-matching} condition~\cite{phm} demands 
\begin{eqnarray}
\phi-\varphi = O(\alpha) \label{phcon}
\end{eqnarray}
for Grover's algorithm to succeed. For large database size, $N \gg 1$ and $\alpha  = 1/\sqrt{N} \ll 1$ and the above condition becomes quite stringent. From the implementation point of view, satisfying Eq. \ref{phcon} is tough as the phase rotations on state $\vert s\rangle$ and $\vert t\rangle$ are not equal in general. Therefore, as the size of the database increases, there is a high risk that Grover's algorithm fails even if there are very small errors in the implementation of the operators.

\begin{figure}[tb]
{\bf (a)}\subfigure{\epsfig{file=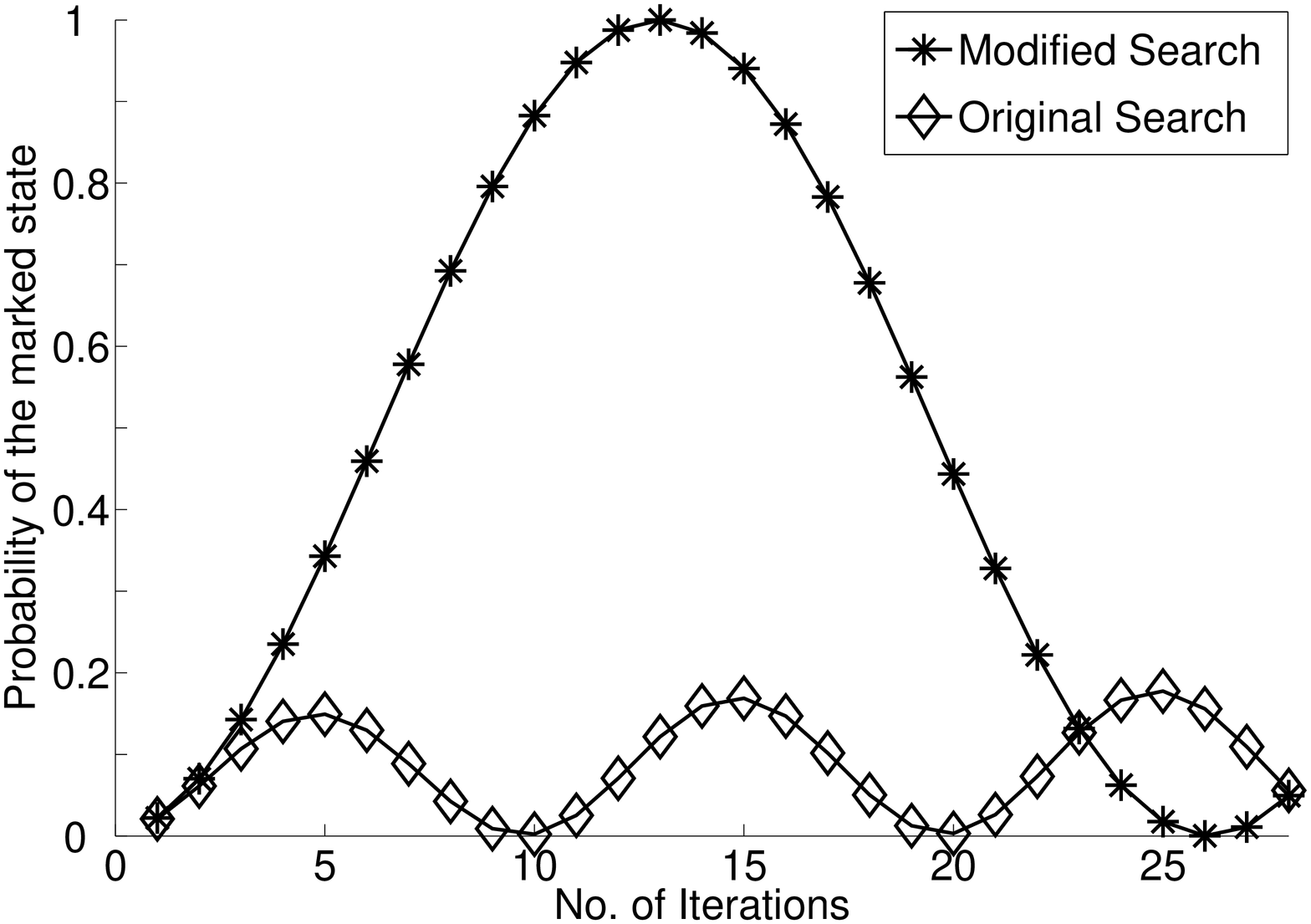,height=0.45\textwidth,width=0.3\textwidth,angle=270} \label{simu}}
{\bf (b)}\subfigure{\epsfig{file=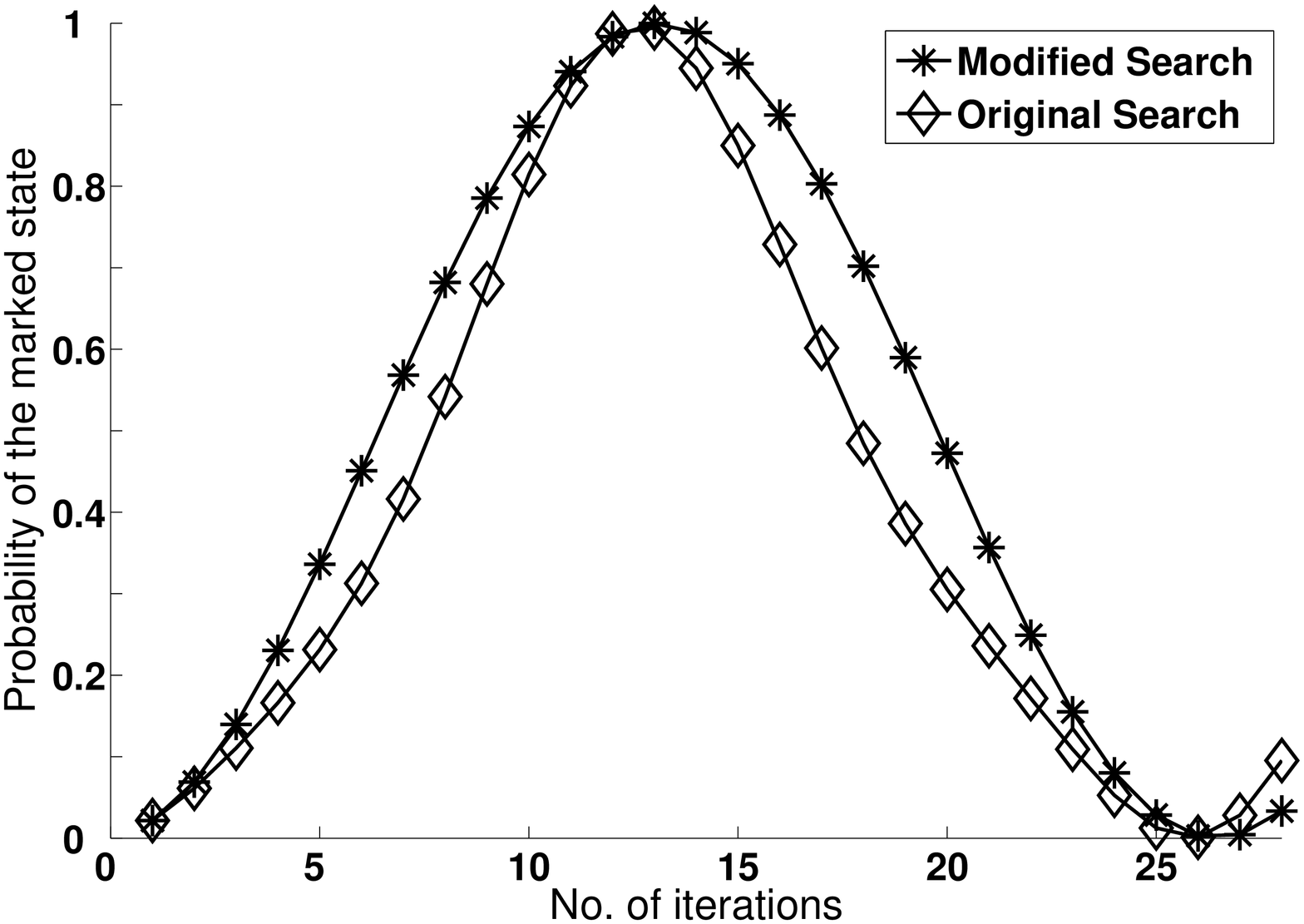,height=0.45\textwidth,width=0.3\textwidth,angle=270} \label{simu1}}
\caption{Simulation results for the original and the modified search algorithm. The line joining the points is just a guide for the eye.
(a) Here $\alpha=0.00091$, $\phi=\pi$ and $\varphi=0.9\pi$. Thus the phase-matching condition $\phi-\varphi = O(\alpha)$ is not met. For the modified algorithm, the probability of the marked state reaches very near to $1$ whereas for the original algorithm, it is always low. (b) Here $\alpha=0.00091$ and $\phi=\varphi=0.9\pi$. Thus the phase-matching condition is satisfied. Grover's algorithm is self correcting even for large errors as long as the phase matching condition is met and its performance is quite similar to that of the modified algorithm.}
\end{figure}

To take into account the above mentioned problem, Tulsi has modified the quantum search algorithm~\cite{tulsi}. The algorithm is based on the assumption that errors are (i) reproducible and (ii) reversible. The reproducibility allows us to implement the transformations $\{I_{s}^{\phi},I_{t}^{\varphi}\}$ at our disposal while the reversibility allows us to implement the inverse transformations $\{I_{s}^{-\phi},I_{t}^{-\varphi}\}$ at our disposal. Then the collective effect of the errors can be cancelled by iterating the following operator
\begin{eqnarray}
\mathcal{T}= I_{s}^{-\phi}I_{t}^{-\varphi}I_{s}^{\phi}I_{t}^{\varphi} = UI_{0}^{-\phi}U^{\dagger}I_{t}^{-\varphi}UI_{0}^{\phi}U^{\dagger}I_{t}^{\varphi} \ . \label{newop}
\end{eqnarray}
Note that for $\phi = \varphi =\pi$, $\mathcal{T}$ is just two steps of Grover's algorithm, i.e. $\mathcal{T}  =\mathcal{G}^{2}$. Tulsi has shown that $\pi/4\alpha \sin \frac{\phi}{2}\sin\frac{\varphi}{2}$ times iteration of $\mathcal{T}$ on $|s\rangle$ yield the target state $|t\rangle$ with high probability. Therefore, if $\alpha$ is small (i.e. the database is large), small difference between $\phi$ and $\varphi$ can cause the Grover's algorithm to fail while the modified algorithm still succeeds in finding the target state (see Fig. \ref{simu} for simulation results). However it may be pointed out that Grover's algorithm is self correcting if $\phi=\varphi$ (Fig. \ref{simu1}). The complexity of both the algorithms remains almost the same for $\{\phi,\varphi\} \not\ll \pi$~\cite{tulsi}. 

It should be noted that for the experimental demonstration of the difference between the original and the modified search algorithm for large database, it is not necessary to implement them on a very large system. We can simulate the behaviour of the algorithms for large database by preparing a small system with $\alpha  = |\langle t|s\rangle| \ll 1$, i.e. initially, the target state has a low probability amplitude. As there are no other restrictions on the size of system, a two qubit system is suitable enough for this purpose.  \\
\indent To experimentally verify the algorithm of Tulsi \cite{tulsi}, the original and the modified search algorithms are implemented here in an NMR Quantum Information Processor. The implementation procedure consists of (i) preparation of the pseudo-pure state (PPS), (ii) preparation of the superposition of all the states such that the marked state has a low probability amplitude, (iii) application of the original/modified iterations and finally (iv) measurement. The experiment has been carried out at room temperature in 11.7 Tesla field in a Bruker AV500 spectrometer using a QXI probe. The system chosen for the implementation of the algorithm is Carbon-13 labeled chloroform ($^{13}$CHCl$_3$), where $^1$H and $^{13}$C act as the two qubits. The $^1$H and $^{13}$C resonance frequencies at this field are 500 MHz and 125 MHz respectively and the scalar coupling between the spins 
is J$_{\mathrm{HC}}$= 209 Hz. The NMR Hamiltonian for a 2-qubit weakly coupled spin system is \cite{ernst},
\begin{eqnarray}
{\mathscr H}= \nu_1 I_{z}^{1} + \nu_2 I_{z}^{2} + J_{12}I_{z}^{1}I_{z}^{2}, \label{nmrham}
\end{eqnarray}
where $\nu_i$ are the Larmour frequencies and the J$_{12}$ is the scalar coupling. The equillibrium density matrix, which is the starting point of any algorithm in NMR quantum information processor, under high temperature and high field approximation is in a highly mixed state represented by \cite{ernst},
\begin{eqnarray}
\rho_{eq}=\gamma_H I_{z}^{H} + \gamma_C I_{z}^{C} = \gamma_H\left(I_{z}^{H} +0.25I_{z}^{C} \right),\label{eq-req}
\end{eqnarray}
where the $\gamma_H:\gamma_C$ is 1:0.25 are the gyromagnetic ratio of the nuclei. We describe the various stages of the experimental implementation in 
the following paragraphs.\\
\begin{figure*}[t]
\epsfig{file=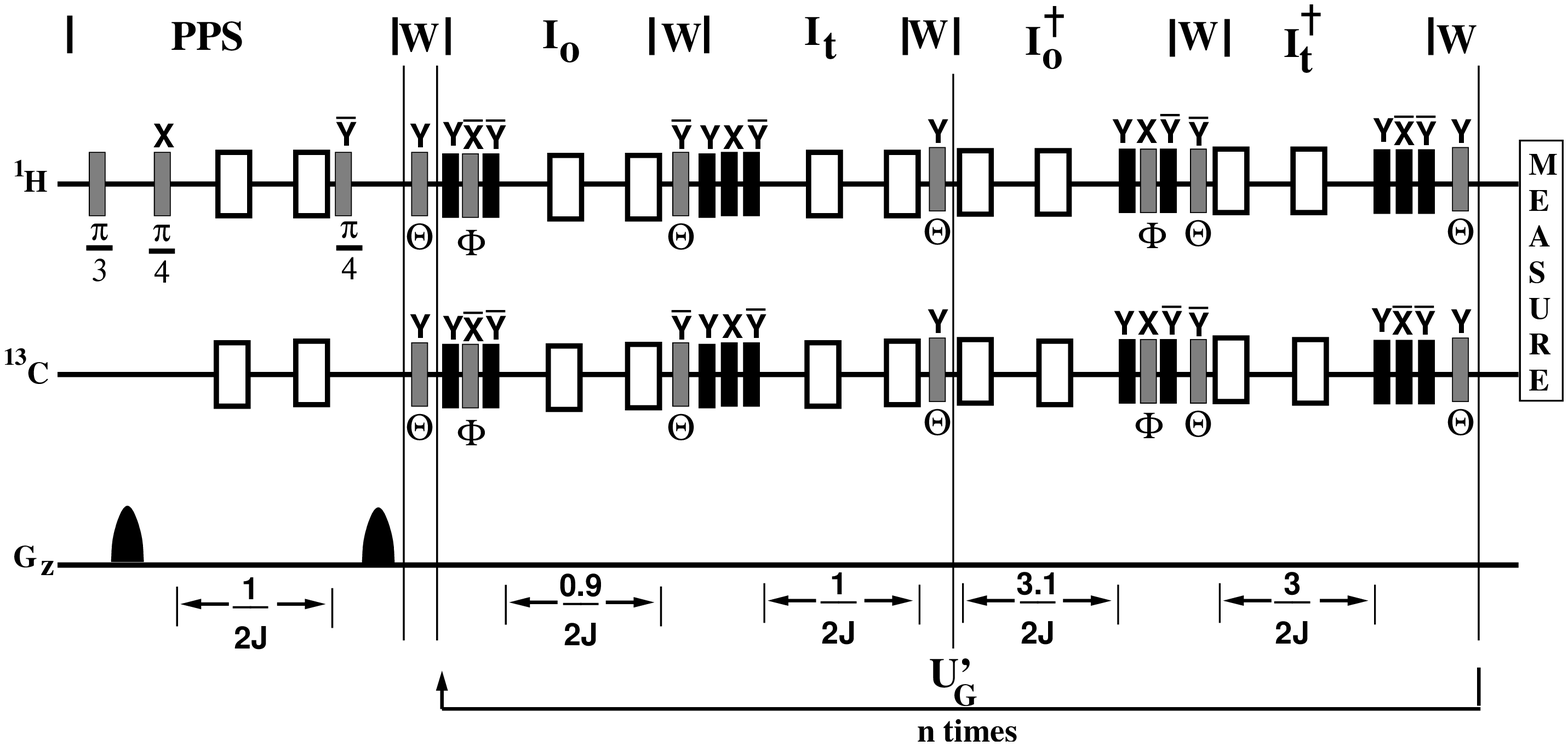,width=0.7\textwidth,height=4cm}
\caption{The pulse sequence for the implementation of PPS and $\mathcal{T}$ operator. The pulses are applied from left to right. All the black narrow pulses have a flip angle of 90$^o$ and the white broad pulse have a flip angle of 180$^o$. The flip angle of the other pulses are written below them. The phases of the pulses are written above them.G$_z$ represent z-gradients which are used to kill unwanted transverse magnetization. The time period of $\frac{1}{2J}$ implies that the system is evolved only under the scalar coupling Hamiltonian and the evolution under the chemical shifts are refocused by the $\pi$ pulses. The flip angle $\Theta=\frac{\pi}{4},\frac{\pi}{6}$ or $\frac{\pi}{9}$ depending upon $\alpha$ and $\Phi=\frac{0.9\pi}{2}$. }\label{pseq}
\end{figure*}
\indent For a two-qubit system, there are $4$ basis states: $|00\rangle$, $|01\rangle$, $|10\rangle$ and $|11\rangle$. We choose the target state to be $|11\rangle$. If $|s\rangle$ is an equal superposition of the basis states then $\alpha = \langle s|11\rangle = 1/2$. But to simulate the Grover's algorithm for large database, we want $\alpha$ to be small. That we achieve by letting $|s\rangle$ to be an unequal superposition. We first create the $\vert 00\rangle$ PPS by the use of spatial averaging \cite{ppsref},[Fig. \ref{pseq}].  A $\vert 00\rangle$ PPS has a unit population in the $\vert 00\rangle$ state and zero population in $\vert 01\rangle, \vert 10\rangle$ and $\vert 11\rangle$ states. Then we apply a $\Theta_y$ pulse on it. We have
\begin{align}
\vert 00\rangle\xrightarrow{\Theta_y}&\left[\cos^2\left(\frac{\Theta}{2}\right)\vert00\rangle+\sin\left(\frac{\Theta}{2}\right)\cos\left(\frac{\Theta}{2}\right)\vert01\rangle\right. 
\cr &\left.+\sin\left(\frac{\Theta}{2}\right)\cos\left(\frac{\Theta}{2}\right)\vert10\rangle+\sin^2\left(\frac{\Theta}{2}\right)\vert11\rangle\right].
\end{align}
Thus $\alpha=\langle s|11\rangle = \sin^2(\frac{\Theta}{2})$. By choosing $\Theta=\frac{\pi}{4},\frac{\pi}{6}$ and $\frac{\pi}{9}$, we achieve $\alpha = 0.146,\ 0.067,\ 0.030$ respectively. Just to compare, if $|s\rangle$ is an equal superposition of $N$ basis states then $\alpha = 1/\sqrt{N}$ and these values of $\alpha$ correspond to $N \approx 47,\  223,\ 1111$ respectively so that we need $n = \log_{2}N \approx 6,8,10$ qubits respectively to represent all basis states. However, by choosing $|s\rangle$ to be an unequal superposition, a two-qubit system becomes sufficient to simulate large databases.


The next step in the implementation of the algorithms is the application of the $\mathcal{G}^2/\mathcal{T}$ operator. In our case, we assume that there are no errors in $I_{t}$ trasnformation, i.e. $\varphi = \pi$. Since $|s\rangle = \Theta_{y}|00\rangle$ and $|t\rangle = |11\rangle$, we have
\begin{eqnarray}
&\mathcal{G}^{2} = \Theta_{y}I_{00}^{\phi}\Theta_{y}^{\dagger}I_{11}\Theta_{y}I_{00}^{\phi}\Theta_{y}^{\dagger}I_{11}\ ,&\\
&\mathcal{T} =  \Theta_{y}I_{00}^{-\phi}\Theta_{y}^{\dagger}I_{11}\Theta_{y}I_{00}^{\phi}\Theta_{y}^{\dagger}I_{11}\ .&
\end{eqnarray}
Note that in case of no errors in $I_{00}$ transformation, we have $\phi = \pi$. Fig. \ref{pseq} contains the pulse programme for the implementation of the $\mathcal{T}$ operator. The $I_{00}^{\phi}$ and $I_{11}$ operators are selective phase rotations of $\vert 00\rangle$ and $\vert 11\rangle$ states respectively i.e.
\begin{eqnarray}
\vert 00\rangle + \vert 01\rangle + \vert 10\rangle + \vert 11\rangle &\xrightarrow{I_{00}^{\phi}}& e^{\imath \phi}\vert 00\rangle + \vert 01\rangle + \vert 10\rangle + 
\vert 11\rangle, \\
\vert 00\rangle + \vert 01\rangle + \vert 10\rangle + \vert 11\rangle &\xrightarrow{I_{11}}& \vert 00\rangle + \vert 01\rangle + \vert 10\rangle - 
\vert 11\rangle.
\end{eqnarray}
Therefore in NMR, the $I_{00}^{\phi}$ and $I_{11}$ operators are implemented by evolution under 
\begin{eqnarray}
I_{00}^\phi\equiv \exp{\left(I_z^1 + I_z^2 + 2I_z^1I_z^2\right)} \label{ioev} \\ 
I_{11}\equiv \exp{\left(-I_z^1 - I_z^2 + 2I_z^1I_z^2\right)} \label{itev}
\end{eqnarray}
respectively. Following \cite{grochu}, for $I_{00}^{\phi}$, the evolution under $I_z^1$ and $I_z^2$ are implemented by composite z-rotation pulses like $\left[\frac{\pi}{2}\right]_{Y}\left[\frac{\phi}{2}\right]_{\bar{X}}\left[\frac{\pi}{2}\right]_{\bar{Y}}$. The evolution under the $2I_z^1I_z^2$ is implemented by evolving the system under the scalar coupling Hamiltonian only, for a time period of $\phi/2\pi J$. The $I_{00}^{-\phi}$ operator is applied by (a) reversing the order of application of pulses and evolution, (b) flipping the phase of the centre pulse of the composite z rotation by $\pi$ and (c) changing the evolution time from $\phi/2\pi J$ to $(4\pi - \phi)/2\pi J$ for $I_{00}^{-\phi}$. The application of $I_{11}$ involves evolution of the system under the Hamiltonian given by Eq. \ref{itev}. This is similar to the Hamiltonian evolution for $I_{00}^{\phi}$, the only difference being the negative sign before $I_z^1$ and $I_z^2$. This implies that the phase of the centre pulse in the composite z rotation is changed by $\pi$. Moreover, as errors are not introduced in the $I_{11}$ operator (i.e. $\varphi = \pi$), the central pulse in the composite z rotation has a flip angle of $\frac{\pi}{2}$ and the time of evolution is 1/2J in both the cases of implementation. \\
\indent After the implementation of the algorithm, the final state is measured. In this case only the diagonal elements of the final density matrix
(population spectra) is required to be measured. This is done by collecting the data after applying a gradient to kill the off-diagonal elements followed 
by a 90$^o$ pulse. The diagonal elements of the final density matrix are reconstructed from the population spectrum.\\
\begin{figure}[t!]
{\rm (a)}\subfigure{\epsfig{file=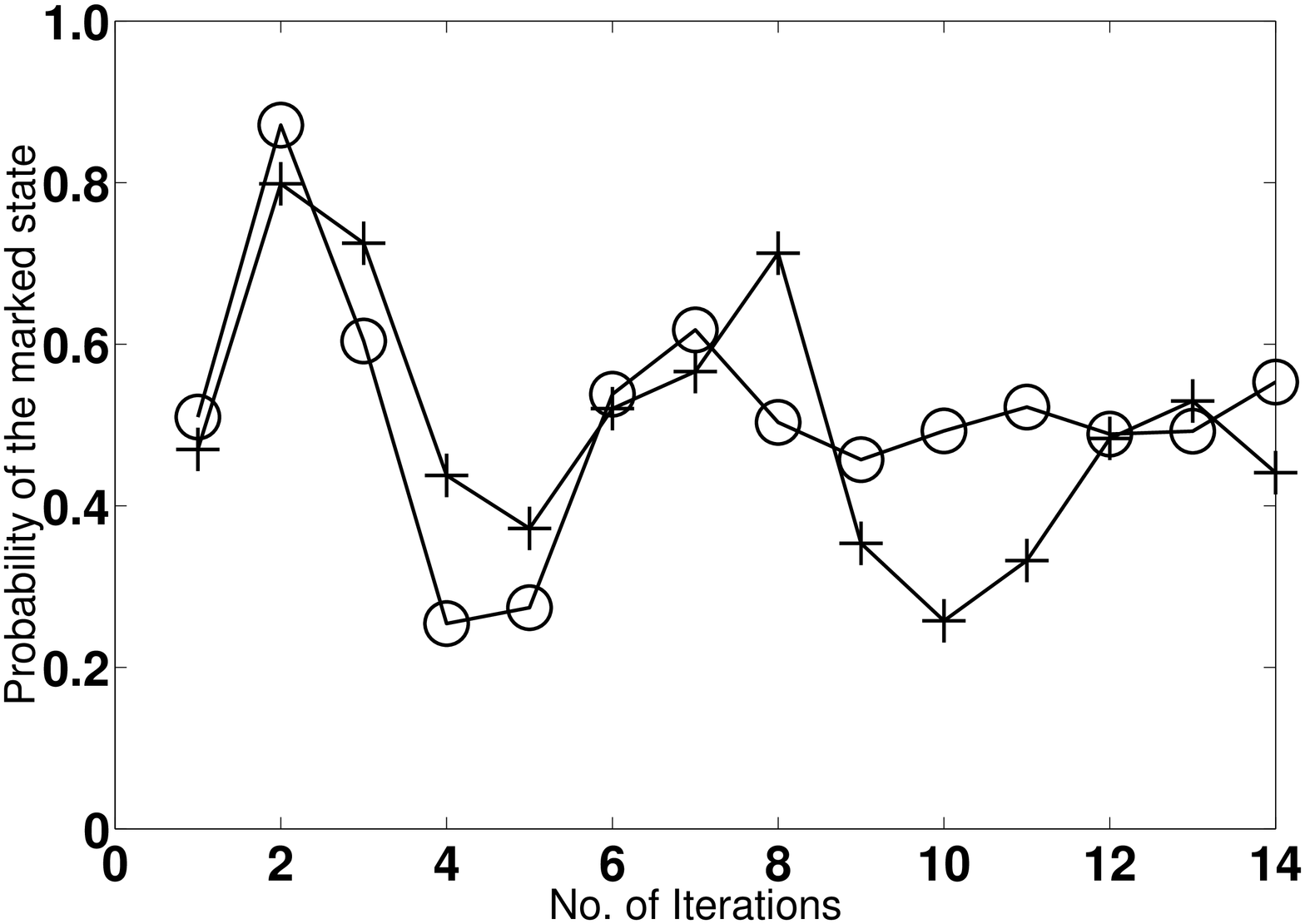,height=0.45\textwidth,angle=270} \label{f3a}}
{\rm (b)}\subfigure{\epsfig{file=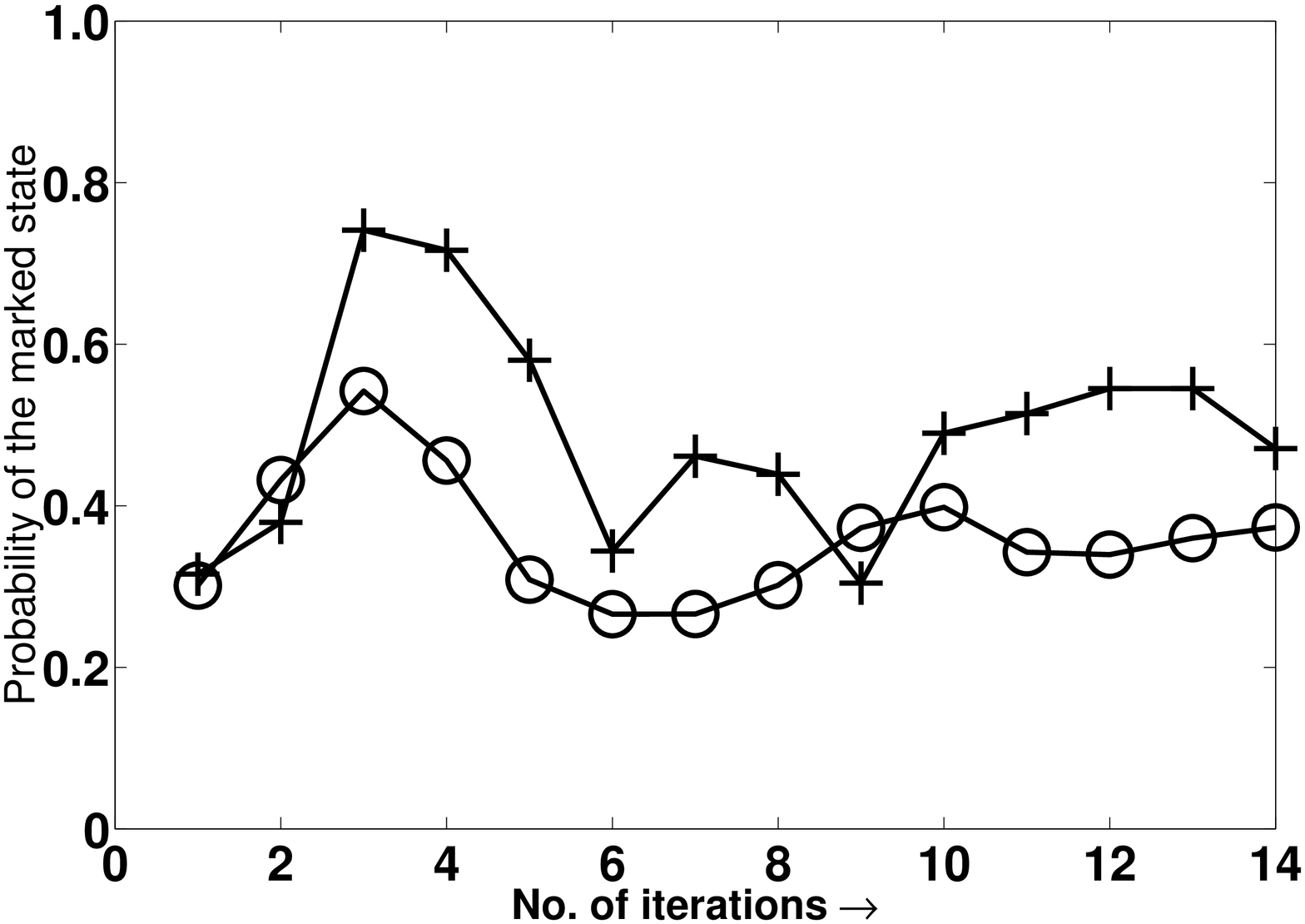,height=0.45\textwidth,angle=270} \label{f3b}}
{\rm (c)}\subfigure{\epsfig{file=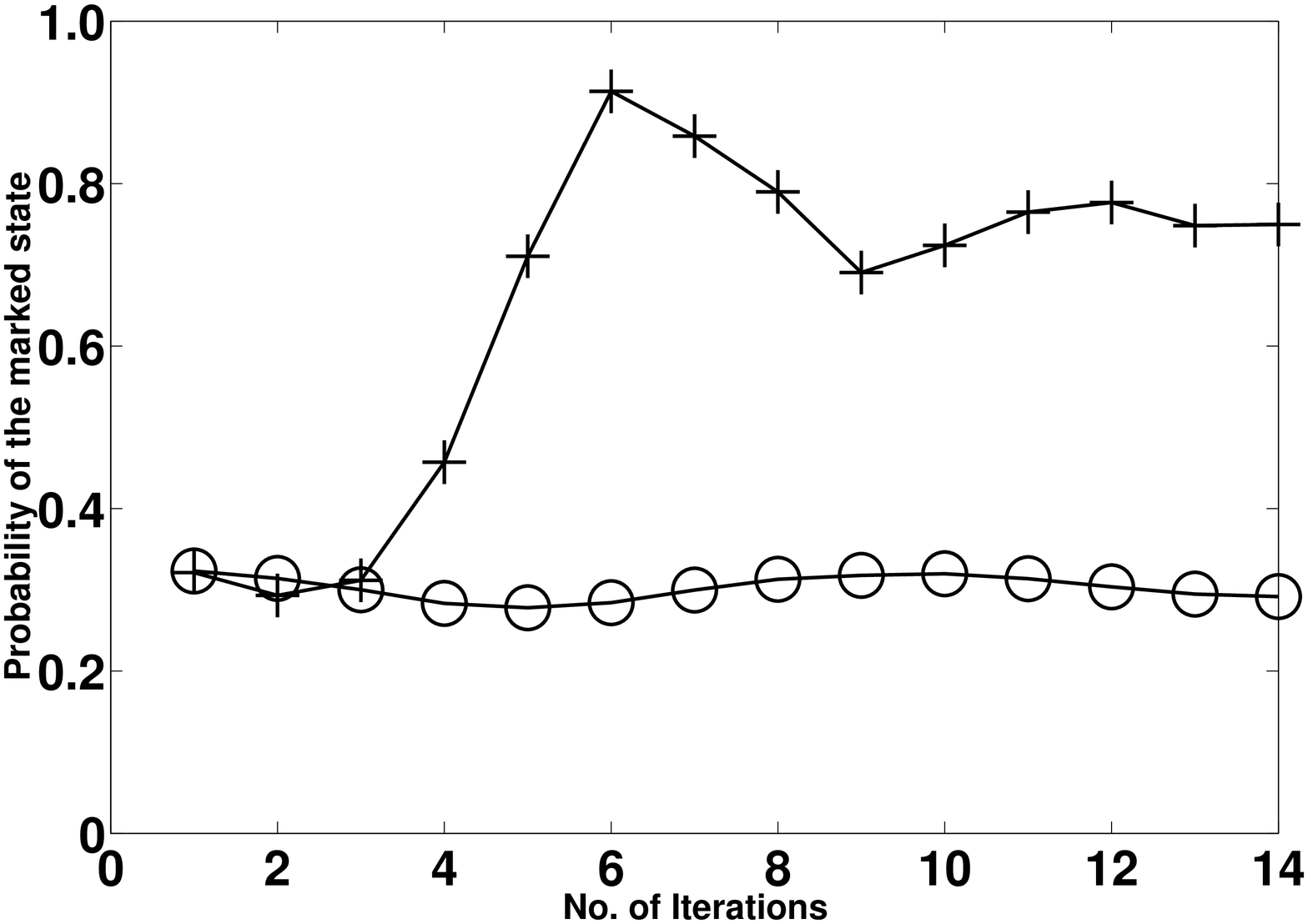,height=0.45\textwidth,angle=270} \label{f3c}}
\caption{The plot of `probability of the marked state' calculated from the measurement of the population (see text) of the final state for each iteration
 of the `original' (depicted by open circles) and the `modified' (depicted by crosses) search algorithm. The line joining the points are a guide to the 
eye. As the value of $\Theta$ decreases $\alpha$ also decreases i.e the size of database increases. (a) $\Theta=\frac{\pi}{4}$ and no error in $I_{s}$ or $I_t$. Both the `original' and the `modified' search algorithm yield the marked state with high probability. (b) $\Theta=\frac{\pi}{6}$ and a 10$\%$ error in $I_s$. The `original' algorithm cannot amplify the amplitude of the marked state to the desired level while the modified algorithm is able to search the marked state with a much higher probebility. (c) $\Theta=\frac{\pi}{9}$ and 10$\%$ error in $I_s$. The `original' algorithm has failed completely in searching the marked state, while the `modified' algorithm succeeds.}
\end{figure}
\indent Fourteen iterations of the original and the modified algorithms were implemented for three different values of initial probability amplitude of the marked state i.e. $\Theta=\frac{\pi}{4},\frac{\pi}{6}$ and $\frac{\pi}{9}$. For the case $\Theta=\frac{\pi}{4}$, no error has been introduced in $I_{00}$ operator (i.e. $\phi = \pi$) which implies that the phase matching condition is satisfied in this case. It can be seen that both `original' and `modified' algorithm behaves almost similarly i.e. they find the marked state with almost the same periodicity (Fig. \ref{f3a}). In the next case (Fig. \ref{f3b}), 
the value of $\Theta$ as chosen to be $\frac{\pi}{6}$ to make $\alpha$ smaller and an error of 10$\%$ was introduced in $I_{00}$ (i.e. $\phi = 0.9\pi$) so that the phase matching condition is violated. We see that in this case, the original search algorithms starts to fail while the `modified' algorithm obtains the searched state with a high probability ($\sim$ 80 $\%$). The original algorithm cannot amplify the amplitude of the marked state so as to definitely distinguish it and therefore the solution is not reached. Finally, the algorithms were implemented for $\Theta=\frac{\pi}{9}$ and 10$\%$ error in $I_{00}$ operator (Fig. \ref{f3c}). In this case $\alpha$ is very small (simulating a system of about 10 qubits), and therefore the `{\it phase matching}' condition is violated even more strongly. It can be seen that in this case, the `original' algorithms totally fails in reaching the solution but the `modified' algorithm succeeds. For completeness, the diagonal elements of the tomographed density matrix for the case of Fig. \ref{f3c} i.e. $\Theta=\frac{\pi}{9}$ and 10$\%$ error in $I_{00}$ are plotted in Fig 4. This confirms the success of the `modified' algorithm of Tulsi \cite{tulsi}.\\
\begin{figure}[t!]
\raisebox{1ex}{\rm (a)}\hspace*{-0.5cm}\subfigure{\epsfig{file=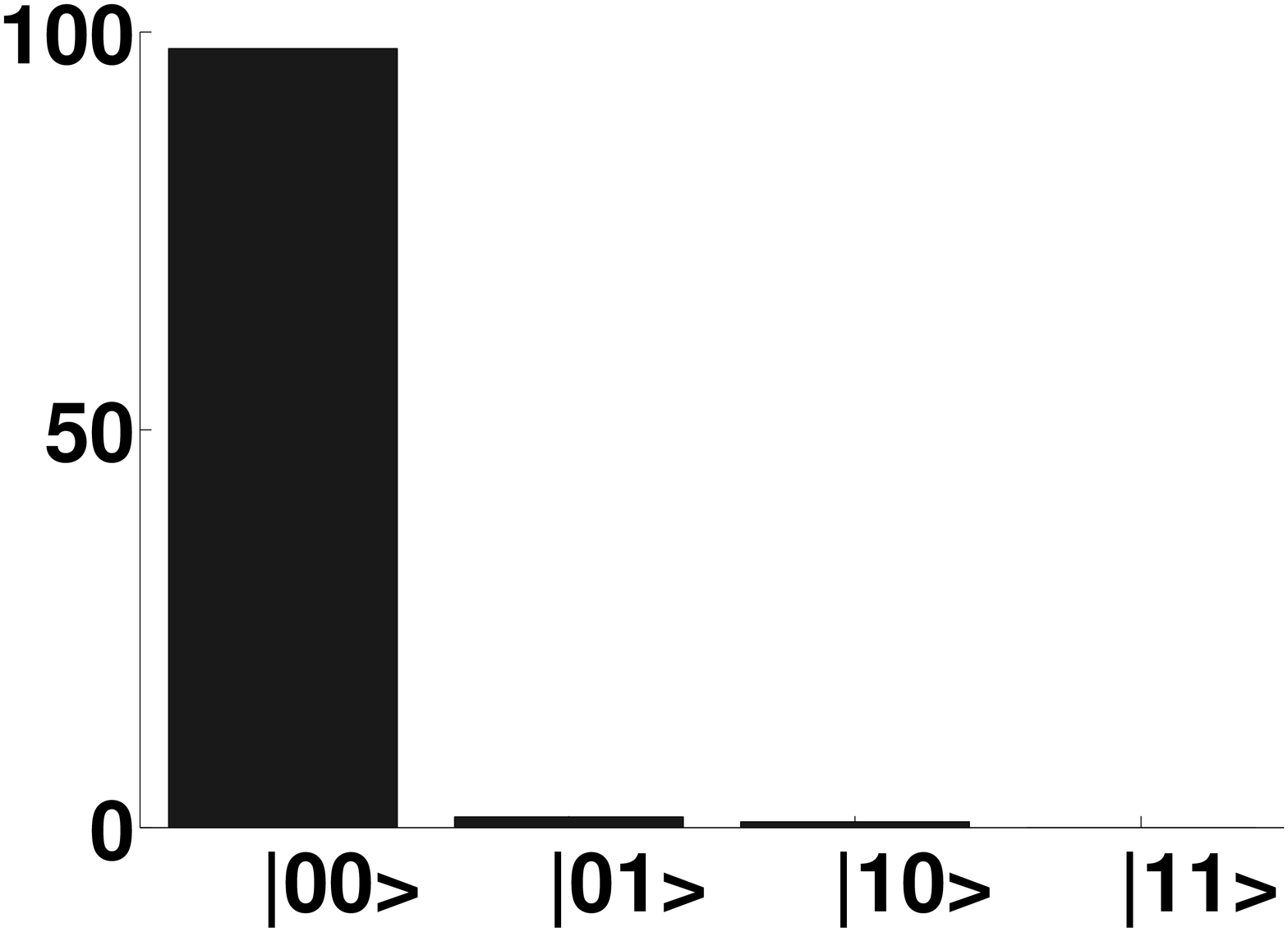,height=0.22\textwidth,angle=270} \label{f4a}} 
\raisebox{1ex}{\rm (b)}\hspace*{-0.5cm}\subfigure{\epsfig{file=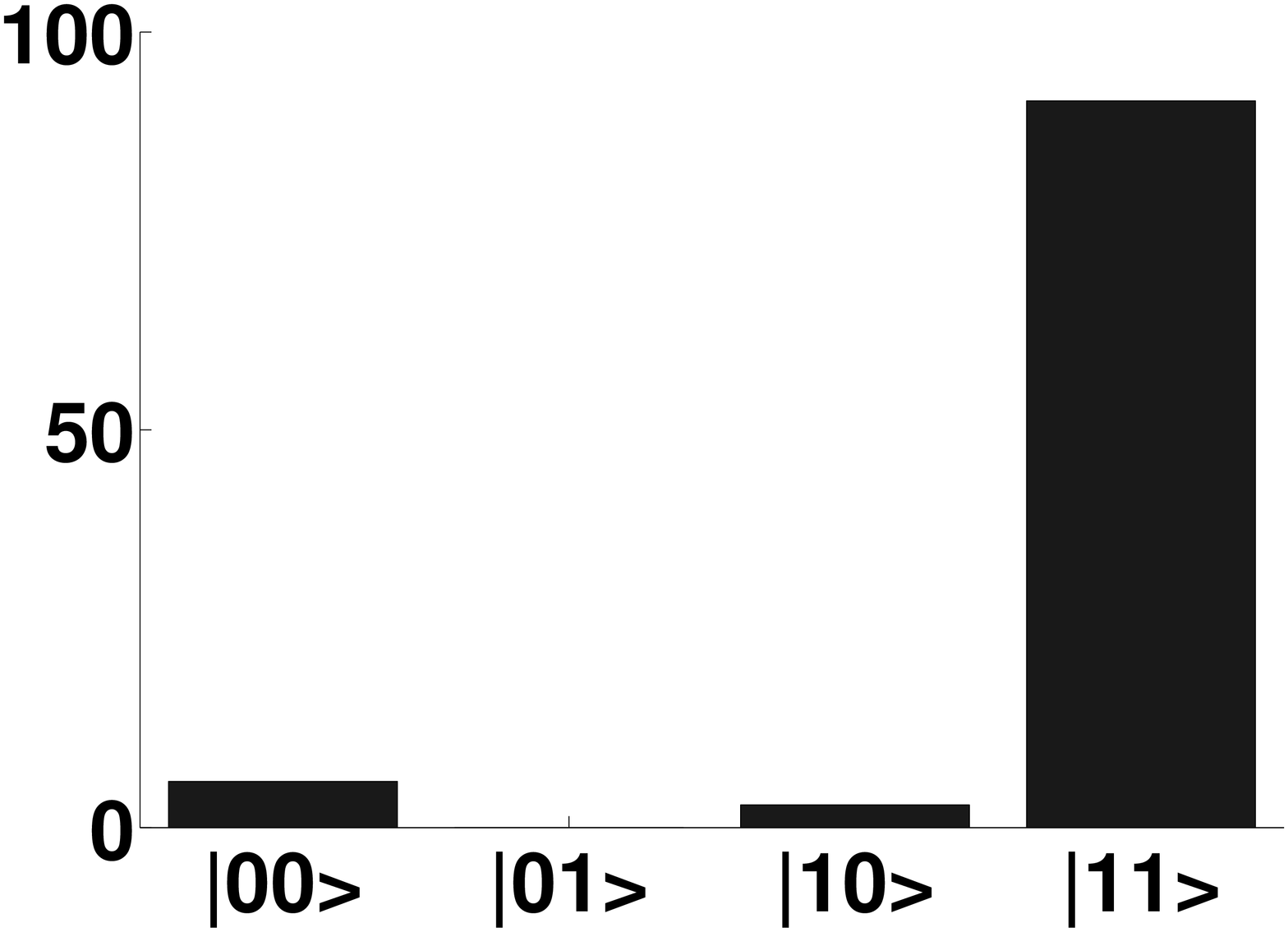,height=0.22\textwidth,angle=270} \label{f4b}}
\caption{ Bar-plot of the diagonal elements of the output density matrix. 
(a) after PPS 
(c) After application of six steps of `modified' search algorithm for $\Theta=\frac{\pi}{9}$ and error in $I_o$ is 10$\%$. 
The marked state $\vert 11\rangle$ has a high probability signifying the sucess of the algorithm. 
}
\end{figure}

\indent In conclusion, we have implemented the `modified' quantum search algorithm by Tulsi \cite{tulsi} and have experimentally validated his claim
that his algorithm is robust to errors in $U_G$ operator as compared to the original search algorithm. We have shown that small errors can be fatal for searching larger databases using Grover's algorithm while the `modified' search algorithm is robust. We have experimentally simulated the behaviour of the algorithms in large database on a 2-qubit NMR quantum information processor. Quantum computers when fully operational will be dealing with real world problems requiring large systems. This experiment, besides providing a validation for an important theoretical prediction, will help in providing impetus to future work on the study of existing algorithms in large real world systems. \\
\uline{Acknowledgment}: The use of AV-500 NMR spectrometer funded by the Department of Science and Technology (DST), New Delhi, at the NMR Research 
Centre, Indian Institute of Science, Bangalore, is gratefully acknowledged. A.K. acknowledges DAE and DST for Raja Ramanna Fellowships, and DST for a research grant on \tql Quantum Computing using NMR techniques\tqr.

\end{document}